\documentclass{article}

\usepackage{microtype}
\usepackage{graphicx}
\usepackage{subcaption}
\usepackage{booktabs} % for professional tables

\usepackage{hyperref}

\usepackage[preprint]{icml2026}

\usepackage{amsmath}
\usepackage{amssymb}
\usepackage{mathtools}
\usepackage{amsthm}

\usepackage[capitalize,noabbrev]{cleveref}

\theoremstyle{plain}

\theoremstyle{definition}

\theoremstyle{remark}

\usepackage[textsize=tiny]{todonotes}

\icmltitlerunning{Systematic LLM Translation of Legacy Scientific Code to Differentiable Frameworks}

\begin{document}

\twocolumn[
  \icmltitle{Systematic LLM Translation of Legacy Scientific Code to Differentiable Frameworks: Application to a Land Surface Model}



  \icmlsetsymbol{equal}{*}

  \begin{icmlauthorlist}
    \icmlauthor{Aya Lahlou}{yyy}
    \icmlauthor{Linnia Hawkins}{yyy}
    \icmlauthor{Pierre Gentine}{yyy}
  \end{icmlauthorlist}

  \icmlaffiliation{yyy}{Department of Earth and Environmental Engineering, Columbia University, New York, United States}

  \icmlcorrespondingauthor{Aya Lahlou}{al4385@columbia.edu}

  \icmlkeywords{differentiable programming, Earth system models, JAX, land surface model, Agentic AI, Code translation, automatic differentiation, parameter calibration}

  \vskip 0.3in
]



\printAffiliationsAndNotice{}  % no special notice (required even if empty)

\begin{abstract}
Differentiable programming offers transformative capabilities for scientific modeling, enabling gradient-based parameter estimation, sensitivity analysis, and data assimilation. Yet, migrating legacy codebases into differentiable frameworks remains a challenge. We present a five-phase LLM-based agentic pipeline that translates legacy Fortran into JAX: static dependency analysis determines module translation order from the full call graph; iterative compile-repair loops correct errors autonomously; and a Fortran reference oracle enforces numerical parity at the module level before integration and gradient verification. We instantiate and evaluate the pipeline on CLM-ml-v2, a 19,000-line Fortran land surface model, and analyze agent behavior across 73 module translation tasks. The resulting differentiable model computes the complete Jacobian in a single backward pass, recovers physical parameters in eight times fewer steps than gradient-free optimization, and achieves a 24 times wall-clock speedup over sequential Fortran at ensemble size N=2,048. Both the translated model and pipeline infrastructure are released as a reusable framework for differentiating other Earth system model components.
\end{abstract}

\section{Introduction}

Legacy Scientific Models, such as Earth system models (ESMs), are among the most consequential yet most outdated large software systems in active use \cite{Neumann2019}. Popular ESMs comprise 500{,}000 to 1.3 million lines of Fortran code~\citep{mendez2014climate, mendez2016legacy}, which support international climate policy and forecast weather for billions of people. Yet, surveys show pervasive deprecated components, poor modularity, missing test coverage~\citep{mendez2014climate}, and a steadily shrinking developer workforce \cite{ecmwf2024modernisation}.

Despite active community efforts to modernize these codebases, no comprehensive, fully differentiable Earth system model exists to date~\citep{gelbrecht2023diff}. End-to-end differentiable models enable gradient-based joint calibration of physical parameters~\citep{raoult2025review, qu2024joint}, automatic adjoint generation for variational data assimilation, replacing decade-long hand-coded adjoint maintenance projects~\citep{ledimet1986, talagrand1987, giering1998taf}, and online training of neural network closures inside the dynamical core, the only configuration shown to yield stable hybrid climate integrations~\citep{kochkov2024neuralgcm, rasp2018deep}. Reverse automatic differentiation computes the gradient of any scalar loss with respect to all p parameters in a single backward pass~\citep{baydin2018autodiff}, collapsing the  $\mathcal{O}(10^4$--$10^5)$ model evaluations per loss required by ensemble Kalman filtering~\citep{evensen2003enkf} or MCMC~\citep{vrugt2009dream} to a single forward-backward pair.

LLM-based code translation is promising but brittle. \citet{pan2024lost} showed that single-pass LLMs achieve low functional correctness even across high-resource language pairs, and Fortran's under-representation in pre-training corpora compounds the difficulty further~\citep{fortran2cpp2024, llmfortran2cpp2025}. Iterative compile-execute-repair cycles are critical for functionally correct output~\citep{fortrankokkos2025, fortran2cpp2024, ibm2023watsonx}. Thus, the open problem is not whether LLMs can translate Fortran, but how to produce a numerically equivalent, fully differentiable translation of a scientific codebase.

We present a five-phase agentic pipeline that addresses this problem through three design principles missing from prior work: (i) static dependency analysis ensuring topologicaly ordered context-aware translation; (ii) modular numerical parity testing using Fortran reference inputs/outputs; and (iii) comprehensive gradient verification. We apply the pipeline to CLM-ml-v2~\citep{bonan2021}, a 19{,}000-line, 102-module multilayer canopy land surface model coupling leaf photosynthesis (FvCB~\citep{farquhar1980}), stomatal conductance (Medlyn~\citep{medlyn2011}), and Harman-Finnigan roughness sublayer turbulence~\citep{harman2007rsl, harman2008rsl} with a Runge-Kutta solver~\citep{Butcher_1964}, producing \texttt{clm-ml-jax}: a fully differentiable and GPU enabled validated re-implementation.

\section{Related Work}
\label{sec:related}

\subsection{Differentiable Programming for Earth System Models}

Code modernization in Earth system modeling has taken two distinct paths. The first is clean rewriting in modern frameworks, most of which aim for end-to-end differentiability \cite{wang2023climaland, hafner2021veros}, but few models to date have achieved it \cite{kochkov2024neuralgcm}.  \cite{kochkov2024neuralgcm, meunier2025fullydifferentiableneuralocean} rebuilds atmospheric dynamics in JAX and is fully differentiable end-to-end. The second path is faithful translation: preserving the physics and parameterizations of an existing operational model. NoahPy \cite{noahpy2026} translates the Noah LSM to PyTorch, \cite{Davenport2026JCM} implements a differentiable version of SPEEDY-based physics \cite{kucharski2013speedy} parametrization in JAX.  

JAX is a predominant framework for differentiable physics-based simulations  \cite{baydin2018autodiff, gelbrecht2023diff, shen2023differentiable, Campagne2023JAXCOSMO, Bezgin2023, jiang2025} enabling composable autodiff, vectorization, and GPU/TPU simulations without manual porting.
\citet{gelbrecht2023diff} identifies automated code translation as a promising path toward making operational ESMs differentiable; this work investigates that direction using a legacy land surface model.

\subsection{Parameter Estimation in Land Surface Models}

Parameter estimation for Earth System Models, and LSMs in particular, has traditionally relied on sampling methods such as Monte Carlo Markov Chain (MCMC) or ensemble Kalman filtering, requiring $\mathcal{O}(10^3$--$10^5)$ model evaluations \cite{vrugt2009dream, evensen2003enkf}.

Gradient-based alternatives, identified as a key emerging avenue by \citet{raoult2025review}, reduce per-step cost to a single forward-backward pass, independent of parameter count.
AdJULES \citep{raoult2016adjules} demonstrated gradient-based LSM calibration via a proprietary adjoint compiler (TAF \cite{giering1998taf}). Still, TAF carries a commercial license (Fastopt GmbH), operates via a Fortran-only source-to-source transformation with no GPU support, and cannot differentiate using iterative solvers (e.g., bisection root-finding, Monin--Obukhov stability iteration) without manually reformulating each solver.
\texttt{clm-ml-jax} removes all three constraints: the translation pipeline is open, and LLM-driven, JAX/XLA compiles gradients natively to GPU, and iterative solvers are differentiated via implicit function theorem adjoints \cite{blondel2022jaxopt}, a capability absent from prior adjoint-based LSM work.

\subsection{LLM-Assisted Scientific Code Translation}

LLM‑assisted translation of legacy Fortran has been demonstrated for particle‑physics codes (CodeScribe \cite{dhruv2024codescribe}), atmospheric HPC kernels \cite{gupta2025kokkos}, and general Fortran‑to‑C++ tasks \cite{ranasinghe2025llmfortran, li2024fortran2cpp}.Translating entire scientific codebases remains difficult to scale; performance degrades when moving beyond isolated functions, largely due to cross-file dependencies and build-system complexity.  
A further challenge is numerical fidelity. Beyond syntax, translated code must reproduce the original physics to high precision, requiring domain-specific testing infrastructure \cite{pietrini2024bridging, ranasinghe2025llmfortran}. Recent benchmarks reveal a significant deficit in evaluating LLMs for scientific code generation: no standard benchmark exists that measures the scientific validity of LLM‑generated code. For Earth system models in particular, \citet{zhou2024chatgpt} shows that ChatGPT can translate isolated land‑surface routines into JAX. Still, their method operates on individual functions without dependency analysis, provides no systematic numerical parity validation against the Fortran reference, and does not verify the correctness of gradients required for scientific deployment.
Unlike general software correctness benchmarks such as SWE‑bench \cite{jimenez2024swebench}, which define correctness via test‑suite execution, scientific translation demands a domain‑specific oracle that enforces numerical parity with the original physical model.

\section{Methodology}
\label{sec:method}

We describe our five-phase workflow for producing \texttt{clm-ml-jax}. The workflow is instantiated here for CLM-ml-v2 but is designed to be transferred to other ESM parts in the future as part of a large refactoring project. A model is suitable for this workflow if it satisfies three conditions: (a) deterministic single-column execution under fixed inputs and physical constants, (b) accessible Fortran source with a standard batch build system, and (c) external linkage against production object files.

\subsection{Phase 1: Scoping and Dependency Analysis}

Naive file-by-file agent prompting fails on coupled codebases because LLMs hallucinate data types for arguments whose types are implicit in \texttt{USE}-statement chains
spanning multiple files~\citep{pietrini2024bridging, li2024fortran2cpp}. We therefore
apply static analysis to map the full codebase dependency structure before any
Translation begins.

We subject the CLM-ml-v2 module dependency graph to a topological sort
\cite{10.1145/368996.369025} to yield a dependency-respecting translation order:
When module $A$ is translated, all modules it imports have already been constructed,
eliminating the hallucinated-type failures that arise in single-file prompting.

The translation order and other static analysis findings are saved in state documents
(Section~\ref{sec:setup}), serving as the agent's initialization context and task
list. This structured decomposition reduces the translation of a 19,000-line, 102-module codebase to a sequence of bounded, context-complete tasks, making autonomous agentic translation tractable at this scale.

\begin{figure}[h!]
  \centering
  \includegraphics[width=\linewidth]{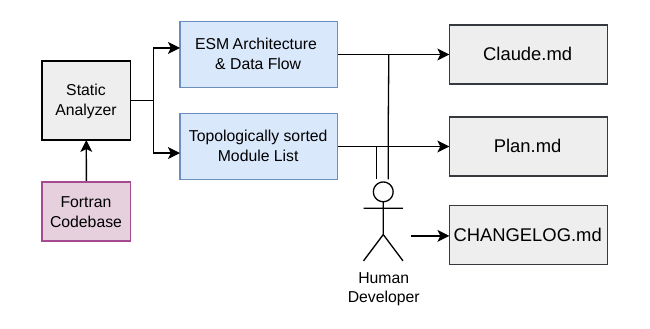}
  \caption{ Flowchart of Static Dependency Analysis and State Documentation Setup.}
  \label{fig:phase1}
\end{figure}

\subsection{Phase 2: State Documentation Setup}
\label{sec:setup}

We establish three persistent state documents checked into the repository (Figure \ref{fig:phase1}): \texttt{CLAUDE.md} encodes project conventions, requirements, and Fortran-to-JAX mapping rules; \texttt{plan.md} tracks per-module, per-phase completion flags; and \texttt{CHANGELOG.md} where agents record current session status and failed attempts. Logging failed attempts is critical for preventing successive sessions from revisiting the same dead-ends. Together, these documents function as external agent memory, preserving project state across context-window boundaries and enabling coherent session resumption without reliance on prior conversation history.

\subsection{Phase 3: Fortran Oracle Construction and Functional Testing}
\label{sec:oracle}
\label{sec:testsuite}

Testing each translated module against its Fortran reference ensures that deviations from scientific ground truth are caught early and attributed to a specific component rather than obscured by the full model's complexity. 

\begin{figure}[h!]
  \centering
  \includegraphics[width=\linewidth]{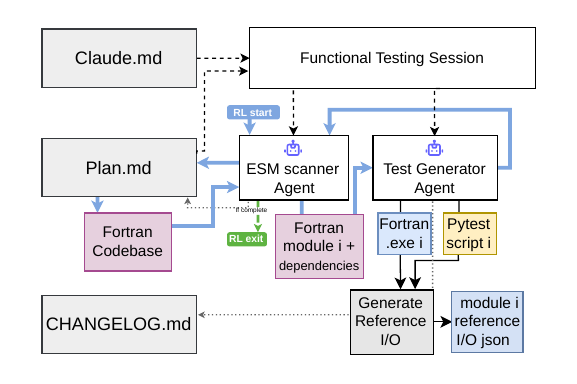}
  \caption{ Fortran Oracle Construction and Functional Testing Flowchart. RL refers to Ralph Loop. Dotted arrows refer to updates to the state documents. }
  \label{fig:phase3}
\end{figure}

We task a team of agents to develop a functional test suite for CLM-ml-v2 following the workflow described in figure ~\ref{fig:phase3}, classifying subroutines into three tiers by testability. Tier~1 subroutines have pure scalar interfaces and are directly callable. Tier~2 subroutines contain isolable physics inside derived-type loops, requiring minor non-breaking visibility changes or scalar helper extraction before testing. Tier~3 subroutines are fully coupled to model state and cannot be tested in isolation; they are instead validated indirectly through full-column parity. This phase produced 26 executables covering tests for 32 subroutines across 10 modules.

The Fortran tests are executed once under fixed inputs, saving the reference input-output
pairs at each checkpoint for validation during the translation phase.

\subsection{Iterative Translation, Testing, and Repair}

\begin{figure}[h!]
  \centering
  \includegraphics[width=\linewidth]{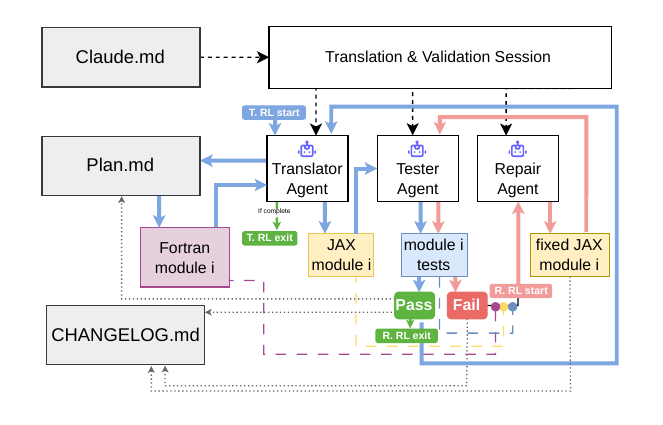}
  \caption{ Autonomous Agentic Translation, Testing, and Repair
Flowchart.  }
  \label{fig:phase4}
\end{figure}

Following the bottom-up dependency order from Phase~1, each module is
processed by two nested autonomous loops (Figure~\ref{fig:phase4}).
A translator agent, guided by \texttt{CLAUDE.md} , produces a JAX equivalent of
the current module. A testing agent then constructs a numerical parity
test, feeding recorded Fortran inputs from the Phase~3 golden I/O files
into the JAX module and asserting that all outputs agree with the Fortran
reference to within a relative tolerance of $10^{-4}$. A passing test
advances the module's \texttt{plan.md} flag and the outer loop moves to
the next module. A failing test engages the inner repair loop: a repair
agent reads the test output, diagnoses the discrepancy, edits the module,
and re-runs the parity test, iterating until the test passes or the module
is flagged in \texttt{CHANGELOG.md} for human inspection.

Both cycles are implemented as \textit{Ralph loops}~\citep{huntley2026ralph}, an orchestration pattern that intercepts the agent's exit signal and re-injects the driving prompt iteratively until a verifiable goal condition is met, enabling unattended multi-hour sessions:

\begin{verbatim}
while true; do
    cat PROMPT_translate.md | claude 
    --dangerously-skip-permissions
done
\end{verbatim}

This design addresses two failure modes of current LLMs in long running
tasks. First, \textit{agentic laziness}\cite{kwa2025laziness}: agents
tend to signal completion prematurely; the loop forces re-evaluation of
all \texttt{plan.md} tasks before exit. Second, \textit{statelessness}:
\texttt{plan.md} and \texttt{CHANGELOG.md} provide the external memory
required for coherent resumption across sessions~\citep{park2023generative}.
Human oversight is maintained asynchronously via \texttt{CHANGELOG.md};
the \texttt{--dangerously-skip-permissions} flag suppresses interactive
prompts for headless operation and should be used only within an
appropriate sandbox. A module passes when its output relative RMSE against the Fortran oracle is below $1\%$ across held-out inputs.

\subsection{Phase 5: Integration and Differentiability Validation}

\begin{figure}[h!]
  \centering
  \includegraphics[width=\linewidth]{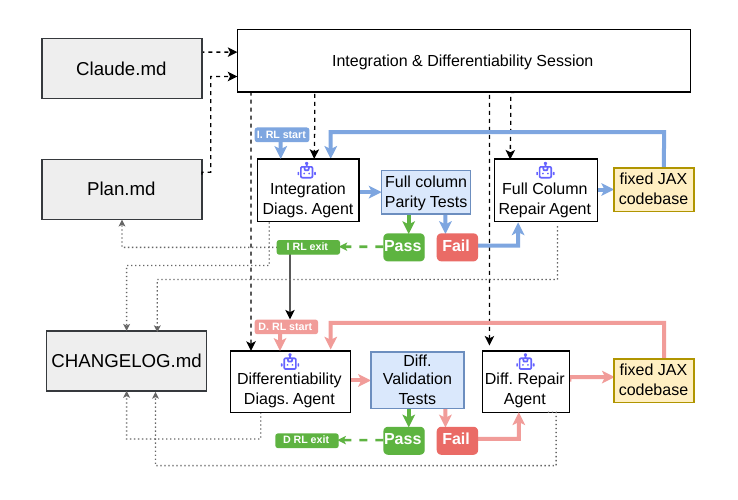}
  \caption{ Integration and Differentiability Validation Flowchart.  }
  \label{fig:phase5}
\end{figure}

Following module-level translation, an Integration Agent assembles the full column pipeline and validates end-to-end behavior against the Full column Fortran oracle as seen in figure \ref{fig:phase5}. This phase revealed errors not exposed by module-level parity tests, such as mismatched array
shapes and indexing conventions across module boundaries. These errors were resolved using the Full Column Repair Agent with shape annotations and systematic indexing checks. The integrated model then achieved full-column parity with the Fortran reference.

Then, a Differentiability Agent is deployed to iteratively identify and resolve JAX-specific issues required for correct and efficient
differentiation. The differentiability validation runs iteratively until all active (output, parameter) gradient pairs fall within 1\% of the finite-difference estimates. Teams targeting other frameworks may encounter analogous issues
(Table~\ref{tab:fortran_patterns} lists the general method and JAX-specific fixes). 

\paragraph{Lessons Learned}from validating the differentiability of \texttt{clm-ml-jax} are as follow:

\textbf{Static loop unrolling} The target framework must trace through the loop body, not execute it iteratively at the Python level. Fortran \texttt{DO} loops over Runge-Kutta sub-steps are replaced with \texttt{jax.lax.scan}, tracing the loop body once at compile time, resulting in a single XLA kernel spanning all sub-steps. This transformation is necessary for efficient gradient computation — JAX can differentiate Python \texttt{for} loops by unrolling them into the computation graph, but doing so for $\mathcal{O}(10^2)$ Runge-Kutta sub-steps creates an unmanageably large computation graph and incurs per-iteration Python--XLA dispatch overhead — and yields a $\sim$200$\times$ reduction in per-step wall-clock time. 

\textbf{Static vs. Dynamic Typing}: Any value that can change between calls must not enter the compile-time cache key. Closures that capture Python scalars extracted from JAX arrays introduce a new value into the XLA cache key on every call, triggering full recompilation ($\sim$290\,s per run). We eliminate this by converting all such constants to JAX scalars before closing over them in \texttt{jit}-compiled functions and caching kernel factories with \texttt{functools.lru\_cache}. After this fix, subsequent runs reuse the compiled kernel ($\sim$0.3\,s cached lookup).

\textbf{Branch-safe Guard} JAX evaluates both branches of a \texttt{jnp.where} expression during the backward pass. Branch-local operations of the form \texttt{x**n} or \texttt{1/x} with $x = 0$ produce \texttt{inf} gradients; the product $0 \times \infty = \text{NaN}$ propagates through the computation graph. We eliminate this by applying safe lower bounds (\texttt{jnp.maximum(x, 1e-30)}) before fractional powers and divisions in five identified modules. 

\textbf{Implicit function theorem (IFT) adjoints for iterative solvers.}Two physics routines implement iterative root-finding via
\texttt{jax.lax.fori\_loop}: the WUE Cowan--Farquhar stomatal conductance
bisection~\citep{cowan1977stomatal} and the Monin--Obukhov length
solver~\citep{monin1954basic}. Differentiating through the loop directly
yields incorrect gradients because \texttt{jnp.where} propagates gradients
through both branches simultaneously at every iteration.
We resolve this using the well-established IFT adjoint
\citep{bai2019deep, blondel2022efficient}: given the converged root
$x^{*}$, we form $x_{\mathrm{IFT}} = x^{*} - F(x^{*};\theta) \;/\;
\texttt{stop\_gradient}(\partial F / \partial x)$, so that in the forward
pass $x_{\mathrm{IFT}} \approx x^{*}$, and in the backward pass
$\partial x_{\mathrm{IFT}} / \partial \theta = (\partial F / \partial
\theta) / (\partial F / \partial x)$. 

\subsection{Agentic Pipeline Characterisation}

To characterize the process for teams applying this methodology to other ESMs, we
tracked all debugging sessions, bug instances, and resolution attempts of Phase~5's differentiability validation
via structured logging in \texttt{CHANGELOG.md}.
Achieving correct end-to-end gradients required fixing 48 documented bugs across 9
mechanistic types over an estimated 46 agentic sessions
(Appendix~\ref{sec:repair_campaign}, Table~\ref{tab:bug_taxonomy}).
The dominant class, \emph{T1 (NaN gradients from \texttt{jnp.where})}, accounts for
35.4\% of bugs (17/48) and follows one mechanical fix pattern:
\texttt{jnp.maximum(x,\,1e{-}30)} applied at 35+ sites across 9 modules; all 17
were resolved on the agent's first attempt. The hardest class, \emph{T5 (gradient
explosion from iterative solvers)}, required an average of 2.0 attempts per bug and
the IFT insight above; without IFT, the Obukhov secant-solver gradient reached
$9.95 \times 10^{144}$. Overall, 81.2\% of bugs were resolved autonomously on the
first attempt; the remaining 18.8\% required human-directed re-specification
(principally T5 and T2 parameter-injection failures, where the failure mode requires
detailed knowledge of JAX's tracing model). Estimated human oversight: $\sim$10--15
hours across 46 agentic sessions. Full taxonomy, timeline, module density map, and
failure analysis are in Appendix~\ref{sec:repair_campaign}.

\section{Experiments}
\label{sec:experiments}

We evaluate clm-ml-jax along two axes: (i) numerical correctness versus the Fortran oracle, and (ii) gradient correctness using central finite differences. We also run proof-of-concept experiments demonstrating the merits of differentiability for parameter tuning in the multilayer canopy land model. All experiments use the CHATS7 walnut orchard AmeriFlux site ~\cite{patton2011chats} with May 2007 meteorological forcing data (30-min resolution). Hardware and software specifications are included in the appendix.

\subsection{Validation of Numerical Equivalence}

\begin{figure}[h!]
  \centering
  \includegraphics[width=\linewidth]{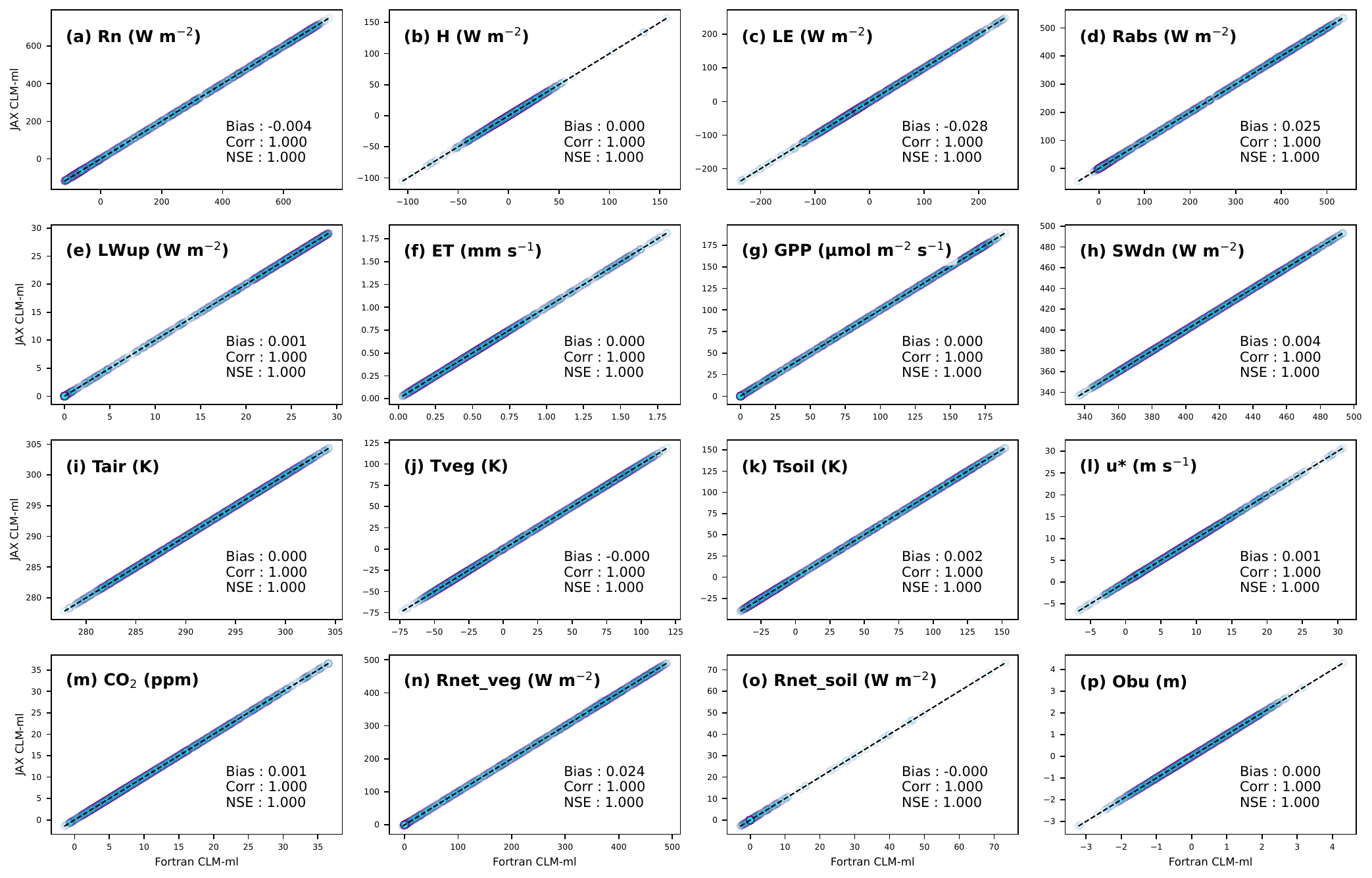}
  \caption{Oracle validation: time series (left) and scatter comparison (right) of JAX vs.\ Fortran outputs for sensible heat (H), latent heat (LE), net radiation (Rn), and GPP across 1488 half-hourly timesteps (May 2007, CHATS7). }
  \label{fig:oracle}
\end{figure}

We compare \texttt{clm-ml-jax} outputs against the Fortran reference across a 31-day simulation in May 2007 at the CHATS7 site. Figure~\ref{fig:oracle} shows time series and scatter comparisons across the full month; canopy profile comparisons are shown in Figure~\ref{fig:profiles}.
\subsection{Validation of Backpropagation Capability}
We validate \verb+jax.grad+ over the CLM-ml-jax column using central finite differences for four key parameters (leaf absorptivity $\alpha_\text{sw}$, $V_{c,\max25}$, stomatal slope, and canopy conductance) and three outputs (GPP, LE, H), using leaf-level fluxes as differentiable proxies (column-level aggregates such as total ET involve non-differentiable accumulation paths in the test harness; leaf-level fluxes isolate the differentiable physics kernel). All four active parameters achieve relative errors less than $1\times10^{-4}$ against central FD for all three output fluxes, well within the 1\% acceptance threshold. A stage-by-stage isolation experiment confirms the gradient path for $\alpha_\text{sw}$: \texttt{d(apar)/d($\alpha_\text{sw}$)} matches FD to $1.2\times10^{-9}$ (solar radiation is exactly differentiable via the Norman two-stream scheme~\cite{norman1979modeling}
), and \texttt{d(agross)/d($\alpha_\text{sw}$)} matches to $1.8\times10^{-7}$ after the implicit function theorem (IFT) fix applied to the stomatal conductance solver. The stomatal conductance  $g_s$ is determined implicitly using a nonlinear equation; direct differentiation through the iterative solver introduces truncation error, so we apply the IFT to obtain exact gradients at the converged fixed point.
The initial four-parameter validation uses the WUE stomatal model; extending to the Medlyn stomatal model~\cite{medlyn2011} with a 10-parameter sweep, 7 of the 10 parameters are structurally active (the remaining 3 are inactive at this stomatal configuration and produce identically zero gradients). All 7 active parameters achieve relative errors less than $1\times10^{-4}$. 

\subsection{Jacobian-Based Sensitivity Analysis}

\begin{figure}[h]
  \centering
  \includegraphics[width=1\linewidth]{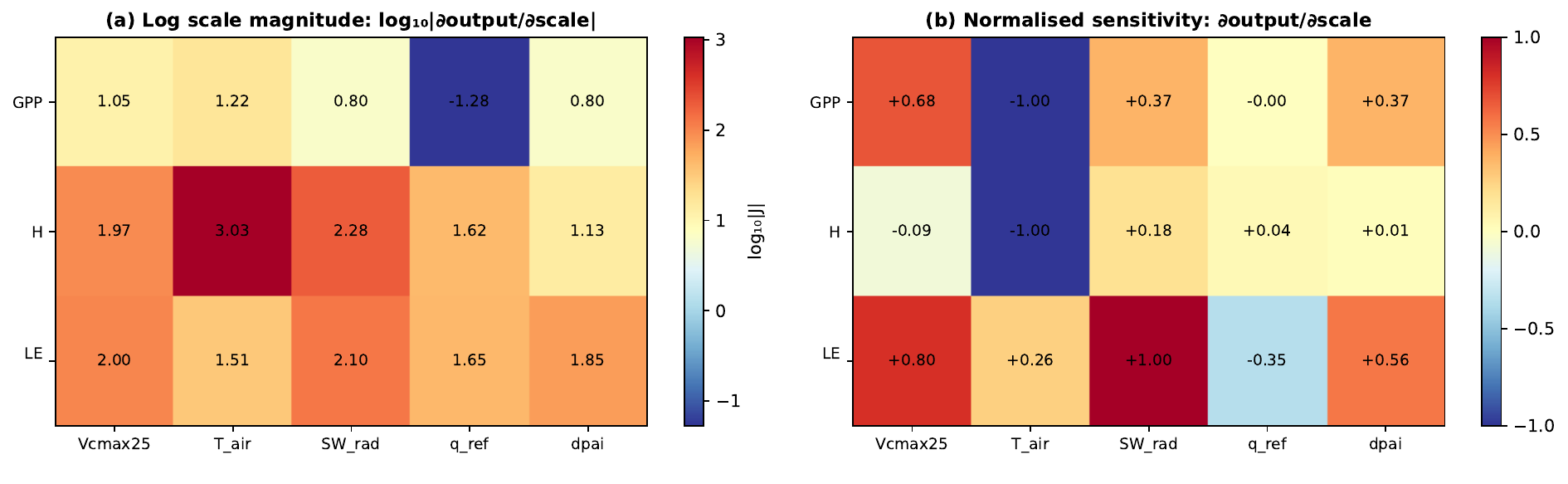}
  \caption{Jacobian-based sensitivity analysis: heatmap of $\partial(\text{GPP},\, H,\,\text{LE})/\partial\boldsymbol{\theta}$ to forcing parameters ($V_{c,\max25}$, air temperature, shortwave radiation, specific humidity $q$, dpai (plant area index per canopy layer)), computed via \texttt{jax.jacrev} in one backward pass. Stomatal model: WUE. GPP, H, and LE are dpai-weighted canopy sums of leaf-level fluxes. \textbf{a)} Log-scale Jacobian magnitude. \textbf{b)} Row-normalized relative sensitivity (each row divided by its maximum absolute value). T\_air dominates H and LE; SW\_rad and $V_{c,\max25}$ are the leading GPP drivers. }
  \label{fig:jacobian}
\end{figure}

We compute $\partial(\text{GPP}, H, \text{LE}) / \partial \boldsymbol{\theta}$ where $\boldsymbol{\theta}$ comprises five scale parameters on: $V_{c,\max25}$, air temperature, shortwave radiation, specific humidity $q$, and plant area index per canopy layer (dpai). GPP is the dpai-weighted sum of gross photosynthesis over canopy layers; H and LE are dpai-weighted sums of sensible heat and latent heat respectively.
$V_{c,\max25}$ is injected via the per-PFT carboxylation capacity dynamic tensor.

\paragraph{Scalar-loss Calibration: Measured AD vs.\ FD}
In the scalar-loss calibration regime ($n=1$ output), the crossover between \texttt{jax.grad} and central FD is governed by the ratio $T_b / T_f$, the cost of one backward pass relative to one forward pass. We measure this directly for $p \in \{1, 2, 3, 5\}$ parameters: the backward pass costs $T_b \approx 4.4\,T_f$ (median across $p$).

\begin{figure}[h]
  \centering
  \includegraphics[width=\linewidth]{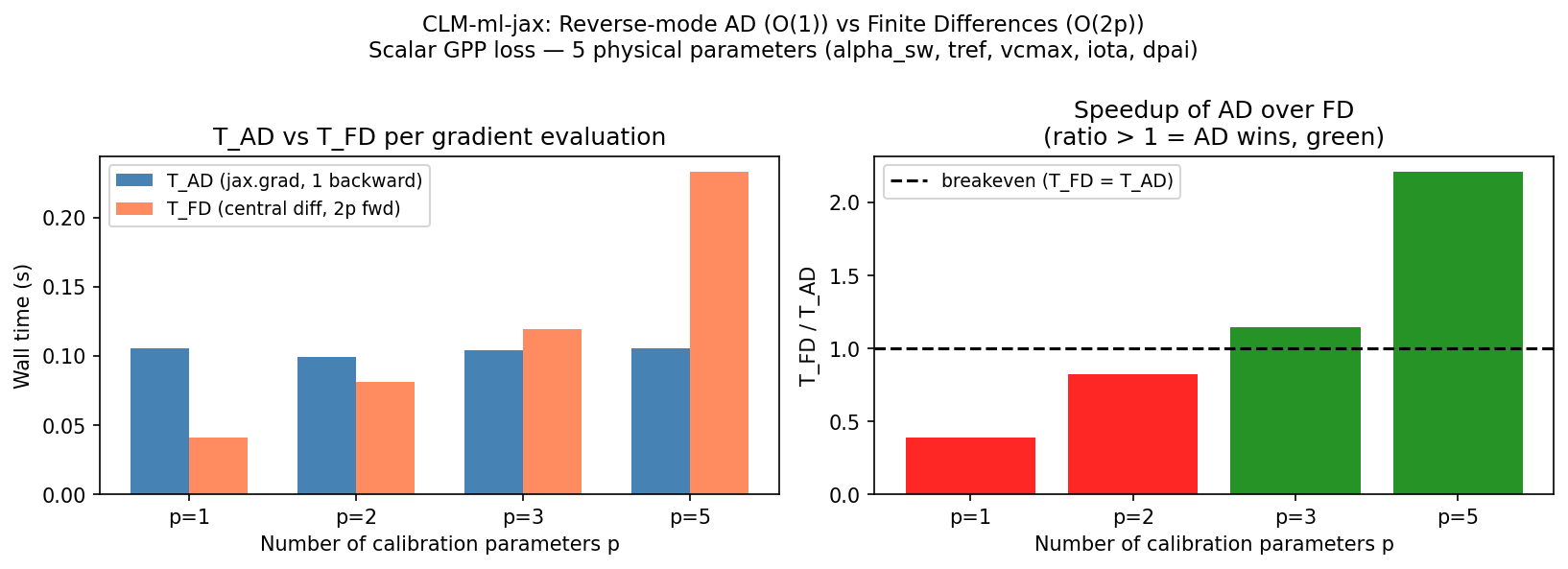}
  \caption{AD vs.\ finite-difference timing as a function of number of parameters $p$ (CHATS7, GPU, scalar GPP loss). Blue bars: $T_\text{AD}$ (\texttt{jax.grad}, constant in $p$). Orange bars: $T_\text{FD}$ ($2p$ forward evaluations, linear in $p$). AD becomes cheaper than FD at $p \geq 3$, confirming the theoretical crossover at $p_\text{cross} = T_b/(2T_f) \approx 2.2$. At $p=5$, AD is $2.2\times$ faster than FD; this advantage grows linearly with $p$.}
  \label{fig:adfd}
\end{figure}

\paragraph{Parameter Calibration - Proof of Concept}

\begin{figure}[h]
  \centering
  \includegraphics[width=\linewidth]{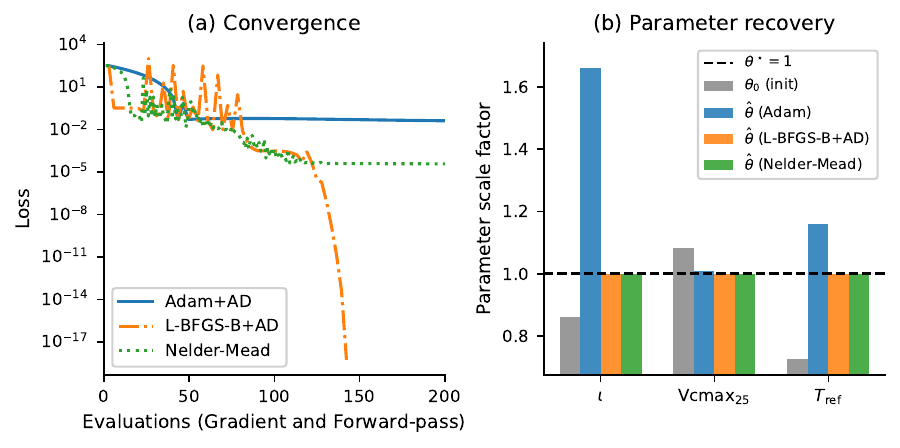}
  \caption{Proof-of-concept parameter recovery for $\iota$, $V_{c,\max25}$, and $T_\text{ref}$ (CHATS7, single timestep, May 2007). \textbf{(a)} Loss vs.\ evaluations (gradient and forward-pass counts combined) for Adam+AD (blue solid), L-BFGS-B+AD (orange dash-dot), and Nelder-Mead (green dotted). L-BFGS-B+AD reaches ${\sim}\,10^{-19}$ in $<\!50$ evaluations; Nelder-Mead requires ${\approx}\,8\times$ more. Adam+AD stalls at ${\sim}\,10^{-2}$ within 100 steps. \textbf{(b)} Parameter recovery ratios $\hat{\theta}/\theta^*$ for all three methods; dashed line at $1.0$ indicates exact recovery ($\theta^*$); grey bars show initial perturbation $\theta_0$.}
  \label{fig:calibration}
\end{figure}

As a proof-of-concept, we calibrate three parameters: stomatal efficiency $\iota$, maximum carboxylation rate $V_{c,\max25}$, and $T_\text{ref}$, against a synthetic target at a single timestep and site (CHATS7, May 2007), bounding the scope of this proof-of-concept to synthetic parameter recovery under controlled conditions. Parameters are initialized from perturbed values; the objective is to recover the known ground-truth $\theta^*$ from the loss landscape. 
We compare three optimizers: L-BFGS-B+AD~\cite{zhu1997algorithm}
, Adam+AD~\cite{kingma2014adam}
(100 cosine-annealed steps), and gradient-free Nelder-Mead~\cite{Nelder1965}
(Figure~\ref{fig:calibration}).

L-BFGS-B+AD reaches machine-precision loss (${\sim}\,10^{-19}$) in fewer than 50 evaluations and recovers all three parameters exactly ($\hat{\theta}/\theta^* = 1$ for $\iota$, $V_{c,\max25}$, and $T_\text{ref}$). Nelder-Mead requires approximately $8\times$ as many evaluations to achieve the same recovery as a gradient-informed quasi-Newton search compared to a gradient-free simplex. Adam+AD stalls at ${\sim}\,10^{-2}$ after 100 cosine-annealed steps, reflecting insufficient iterations for a first-order method on this landscape rather than a failure of differentiability; L-BFGS-B exploits exact curvature information and converges ${\sim}\,17$ orders of magnitude lower within the same evaluation budget.
This efficient gradient-based calibration --- previously unavailable in CLM-ml-v2 --- is the direct scientific payoff of the five-phase agentic translation pipeline.

\subsection{Performance Benchmark}
\label{sec:exp5}

We characterize the wall-clock cost of \texttt{jax.vmap} over parameter ensembles relative to sequential Fortran execution, measuring throughput as a function of ensemble size $N$ on an  Quadro RTX 8000 GPU and sequential CPU, with hardware and software specifications in Appendix~\ref{sec:hardware}.

\begin{figure}[h]
  \centering
  \includegraphics[width=\linewidth]{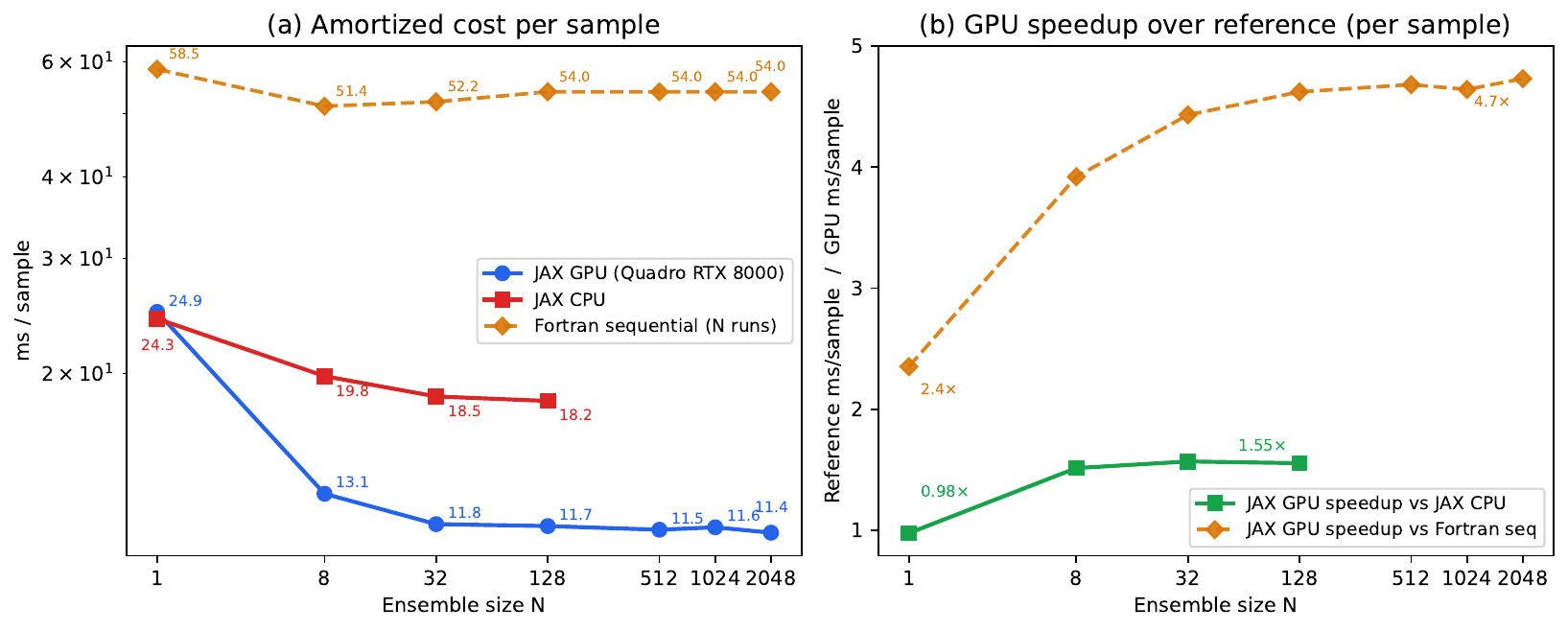}
  \caption{Throughput scaling of CLM-ML-JAX with  
  ensemble size $N$ (CHATS7 walnut orchard, 46 layers,      
  Quadro RTX 8000). $N$ is the number of independent forward
   passes---each with a distinct parameter vector---run
  simultaneously, as needed for ensemble calibration or     
  uncertainty quantification. \textbf{(a)} GPU amortized  
  cost falls from 24.9 to 11.4\,ms/sample as $N$ grows from
  1 to 2048; Fortran executes serially and stays flat at
  ${\approx}$54\,ms/sample. \textbf{(b)} Per-sample GPU
  speedup over Fortran reaches $4.7{\times}$ at $N=2048$ and
   continues to grow with $N$.}
  \label{fig:benchmark}
\end{figure}

\paragraph{Throughput scaling.}
Figure~\ref{fig:benchmark} shows amortized cost per sample as a function
of $N$. GPU cost falls from $24.9$\,ms/sample at $N=1$ to $11.4$\,ms at
$N=2{,}048$, plateauing at $4.7\times$ below the Fortran sequential cost
of ${\approx}54$\,ms/sample. At $N=2{,}048$, Fortran sequential requires
$553$\,s while JAX on GPU requires $23.4$\,s, a ${24\times}$
total wall-clock reduction. This makes multi-start ensemble calibration
and parameter uncertainty quantification tractable at the single-site
scale.

\paragraph{Scan dispatch overhead.}
Replacing the Python-level Runge-Kutta loop with \texttt{jax.lax.scan}
traces the loop body once at compile time, eliminating per-iteration
Python--XLA dispatch overhead. This yields a $164\times$ speedup for
Euler timestepping and a $3{,}100\times$ speedup for fourth-order RK4.

\paragraph{Scalar-loss calibration: AD vs.\ finite differences.}
In the scalar-loss regime ($n=1$ output), the crossover between
\texttt{jax.grad} and central finite differences is governed by
$T_b/T_f$, the cost of one backward pass relative to one forward pass.
We measure $T_b \approx 4.4\,T_f$ (median across $p$; 
Figure~\ref{fig:adfd}). AD becomes cheaper than finite differences at
$p \geq 3$, confirming the theoretical crossover at
$p_{\mathrm{cross}} = T_b / (2T_f) \approx 2.2$. At $p=5$, AD is
$2.2\times$ faster than finite differences; this advantage grows linearly
with $p$.

\paragraph{Numerical precision.}
The model is dominated by memory-bandwidth-bound operations
(\texttt{exp}/\texttt{log}/\texttt{sqrt} in stomatal and radiative
transfer kernels); \texttt{float32} provides no throughput benefit over \texttt{float64} on modern GPUs (${\leq}5\%$ difference across all $N$; Table~\ref{tab:precision}). 

\paragraph{Compile-time scaling.} Applying JIT-compiled \texttt{vmap} execution on CPU requires XLA
to materialise a flat $\mathcal{O}(N \times M)$ LLVM compilation graph
at trace time, where $M$ is the number of model operations; at $N=128$
this exhausts virtual address space regardless of available physical RAM,
as the latter affects only the time to failure, not the outcome. On GPU,
\texttt{vmap} tiles across the batch dimension rather than unrolling it,
keeping compile time $\mathcal{O}(M)$ independently of $N$. GPU execution
is therefore a hard requirement for ensemble sizes $N > 1$, not merely
a performance optimization.

\section{Discussion}
\subsection{Scientific Implications}
The oracle validation establishes that \texttt{clm-ml-jax} is a scientifically equivalent re-implementation of the Fortran model, preserving all PFT parameterizations without recalibration. Verified reverse-mode gradients through the full 46-layer coupled physics column—encompassing IFT solvers, a Runge-Kutta sub-step loop, and tightly coupled leaf-canopy-turbulence physics—confirm that autodiff is accurate through non-trivial numerical structures. The Jacobian analysis yields a concrete scientific result: air temperature and shortwave radiation dominate sensitivities of GPP and energy fluxes at the CHATS site, with $V_{c,\max25}$ exerting moderate but distinct control over GPP. Full-column parameter sensitivities, previously intractable in a single backward pass, directly inform which observations most constrain model predictions at a given site. End-to-end differentiability further positions \texttt{clm-ml-jax} for hybrid physics-ML architectures in which neural parameterizations of uncertain processes are jointly trained with the physical column \cite{aboelyazeed2023, shen2023differentiable}.
\subsection{Limitations and Future Work}
Headless operation requires skipping permissions; operators should apply sandboxing and review \texttt{CHANGELOG.md} before deployment. The calibration experiment remains a proof-of-concept: 3 parameters, a single site, and a single timestep.
Future work includes coupling \texttt{clm-ml-jax} to JAX-based atmospheric models \cite{hafner2021veros} for fully differentiable land--atmosphere simulation, applying the translation methodology to CLM5 and the CLUBB convection scheme in CAM, and training neural parameterizations end-to-end within the differentiable column.
\section{Conclusion}

We have proven three things. First, that a 19,000-line validated Fortran land surface model can be translated to a numerically equivalent, fully differentiable JAX implementation via a five-phase agentic pipeline without manual reimplementation and with confirmed numerical equivalence. Second, we validated the model's end-to-end differentiability: gradients pass finite-difference validation through the full coupled physics column, including IFT-corrected iterative solvers, and the Jacobian recovers physically interpretable sensitivities in a single backward pass. Third, gradient-based parameter optimization (L-BFGS-B) recovers three land surface parameters to machine precision in fewer than 50 evaluations  8$\times$ faster than gradient-free search on the same problem.

Beyond the model itself, this paper is a case study in AI as a tool for scientific code modernization. We present a five-step code translation workflow with autonomous multi-day repair loops, domain-specific oracle validation, and a version-controlled audit trail. The 81.2\% first-attempt autonomous resolution rate and $\sim$10--15 hours of total human oversight across 46 sessions quantify what co-authorship looks like in practice for legacy scientific code. We release the model and the methodology infrastructure to support the community in applying this approach to other ESM components.

\section*{Acknowledgements}
We acknowledge funding from NSF through the Learning the Earth with Artificial intelligence and Physics (LEAP) Science and Technology Center (STC) (Award \#2019625).

\section*{Impact Statement}
Differentiable land surface models enable gradient-based parameter estimation and sensitivity analysis that were previously intractable, with direct relevance to reducing uncertainty in land-atmosphere carbon and energy flux projections. The calibration experiment presented here is a proof-of-concept at single-site, single-timestep scale; extension to multi-site, long-period calibration could meaningfully reduce projection uncertainty in carbon cycle feedbacks.
The agentic pipeline introduces governance considerations specific to autonomous scientific software development. As agentic pipelines assume roles traditionally held by domain scientists, the field requires audit trails and validation standards commensurate with the scientific trust placed in the resulting artifacts. The state-document infrastructure presented here represents a step toward reproducible and auditable agentic workflows in scientific computing.

\bibliography{references}
\bibliographystyle{icml2026}

%%%%%%%%%%%%%%%%%%%%%%%%%%%%%%%%%%%%%%%%%%%%%%%%%%%%%%%%%%%%%%%%%%%%%%%%%%%%%%%
%%%%%%%%%%%%%%%%%%%%%%%%%%%%%%%%%%%%%%%%%%%%%%%%%%%%%%%%%%%%%%%%%%%%%%%%%%%%%%%
% APPENDIX
%%%%%%%%%%%%%%%%%%%%%%%%%%%%%%%%%%%%%%%%%%%%%%%%%%%%%%%%%%%%%%%%%%%%%%%%%%%%%%%
%%%%%%%%%%%%%%%%%%%%%%%%%%%%%%%%%%%%%%%%%%%%%%%%%%%%%%%%%%%%%%%%%%%%%%%%%%%%%%%
\newpage
\appendix
\onecolumn

\section{Hardware and Software Specifications}
\label{sec:hardware}

Performance Benchmarking experiments were run on NVIDIA A40 GPU and CPU nodes. Fortran reference timing measured on NCAR Derecho (Intel Ice Lake CPUs). All other experiments were run on an NVIDIA A100-PCIE-40GB GPU—software: JAX 0.9.2, Python 3.11, CUDA 12.8, gfortran 12.

\begin{table}[h!]
  \caption{Common Fortran patterns encountered in CLM-ml-v2, the general differentiable-programming concept each requires, and the JAX-specific equivalent. The \emph{General concept} column is framework-agnostic; teams targeting PyTorch, Julia Flux, or other frameworks can apply the same concept using their framework's primitives.}
  \label{tab:fortran_patterns}
  \centering
  \small
  \begin{tabular}{ p{.22\linewidth} p{.22\linewidth} p{.22\linewidth} p{.22\linewidth} }
    \toprule
    Fortran Pattern & General concept & JAX Equivalent & Notes \\
    \midrule
    \texttt{DO} loop over sub-steps
      & Static loop unrolling for gradient tracing
      & \verb|jax.lax.scan|
      & Python \texttt{for} loops produce unrolled graphs; \texttt{lax.scan} traces once \\
    \midrule
    In-place array mutation
      & Functional (copy-on-write) update
      & \texttt{x.at[i].set(v)}
      & JAX arrays are immutable; all target frameworks require explicit copy \\
    \midrule
    \texttt{COMMON} blocks / module state
      & Explicit argument threading
      & Explicit argument passing; \texttt{NamedTuple} state
      & No global mutable state in any differentiable framework \\
    \midrule
    \texttt{IF/ELSE} with loop-carried deps
      & Differentiable conditional
      & \texttt{jax.lax.cond} / \texttt{jax.lax.switch}
      & Avoids Python-level branching in JIT-compiled code \\
    \midrule
    Implicit typing
      & Explicit dtype annotation
      & \texttt{jnp.float64} throughout
      & Type consistency required for correct gradient accumulation \\
    \midrule
    Conditional branch with \texttt{x**n} or \texttt{1/x}, $x=0$
      & Branch-safe guard
      & \texttt{jnp.maximum(x,1e-30)} before power/division
      & Both branches evaluated in backward pass; $0 \times \infty = \text{NaN}$ \\
    \midrule
    Iterative root-finding (\texttt{DO} convergence loop)
      & Custom adjoint via implicit differentiation
      & Newton-refinement IFT wrapper
      & Loop-through-convergence gradients incorrect; IFT gives exact adjoint \\
    \midrule
    Module-level mutable arrays
      & Immutable state container
      & NamedTuple with \texttt{.\_replace(field=value)}
      & Enables functional differentiation; PyTorch: \texttt{dataclass}; Julia: struct \\
    \bottomrule
  \end{tabular}
\end{table}

%%─────────────────────────────────────────────────────────────────────────────

\section{Static Analysis: CLM-ml-v2}
\label{sec:static_results_full}

\subsection{Dependency Graph Analysis}
The inter-module \texttt{USE} dependency structure resolves to a directed graph $G_M$ of 76 nodes and 315 directed edges. The graph density is $\rho = 315 / (76 \times 75) = 0.0553$, indicating moderately sparse but non-trivial inter-module coupling, with an average of 8.29 \texttt{USE} references per module. $G_M$ is acyclic: the absence of circular dependencies yields a strict directed acyclic graph (DAG), which is a necessary precondition for safe topological translation ordering.

\begin{figure}[h]
  \centering
  \includegraphics[width=1\linewidth]{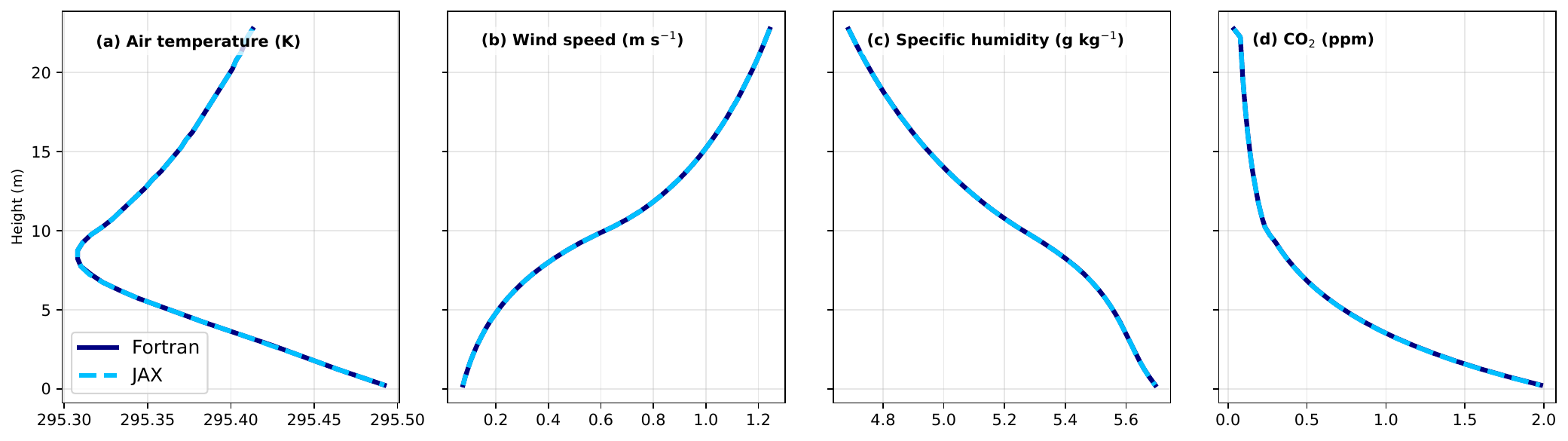}
  \caption{CHATS7 May 1, 2007: Canopy profiles at noon (timestep 24): JAX vs.\ Fortran reference across all 46 canopy layers. Variables shown include air temperature, wind speed, specific humidity, CO$_2$ concentration, and leaf-level photosynthesis and stomatal conductance.}
  \label{fig:profiles}
\end{figure}

\subsection{Implications for Translation Ordering and LLM Context Budgeting}

The strict DAG structure directly enables a topologically safe translation order. Modules are processed in increasing depth order, ensuring every dependency of a given module is translated, with its numerical parity verified against Fortran golden data, before that module's own LLM translation prompt is constructed. The 12 depth levels partition all 73 modules into parallel batches. The maximum achievable parallelism is 15 concurrent translations at depth~2.

\paragraph{Context budgeting.}
Module size varies considerably across the codebase, producing a wide range of per-call token demands. Deep-stack modules require loading the signatures of all previously translated dependencies alongside the source itself. \texttt{MLCanopyFluxesMod.F90} is representative of this challenge: at dependency depth~7 with 40 declared internal dependencies, a single translation call must incorporate up to 40 JAX module signatures in addition to a 1{,}511-line source file.
Static analysis identifies high-effort modules which receive targeted treatment: reference I/O data are generated across physically diverse atmospheric forcing conditions for robust numerical validation, and a higher-capacity model (Claude Opus 4.6) is allocated for agentic translation and repair.

%% ── Repair Campaign Analysis ─────────────────────────────────────────────────
\section{Characterization of Agentic Differentiability Repair}
\label{sec:repair_campaign}
We report statistics of the Differentiability Optimization phase (Phase~5, Part~2), derived from \texttt{CHANGELOG.md}. These statistics characterize the repair effort a team may encounter when applying this methodology to a comparably scaled ESM.
\paragraph{Bug taxonomy.}
Table~\ref{tab:bug_taxonomy} classify 48 documented bugs into 9 mechanistic types. Types T1 and T2 jointly account for 54.2% of all bugs.
T1 bugs follow a single recurring template: inserting \texttt{jnp.maximum(x,1e{-}30)} prior to any division or fractional power within a guarded branch. This fix was applied at 35 or more sites across 9 modules.
T2 bugs decompose into four sub-patterns. (a)~Python \texttt{float()} casts silently break the JAX autodiff tape by detaching values from the trace. (b)~Physics state is overwritten by \texttt{MLCanopyFluxes.\_\_init\_\_} before the traced computation executes. (c)~Functions decorated with \texttt{@jax.jit} capture module-global parameter values as XLA compile-time constants, rendering subsequent module-level mutations invisible to JAX. (d)~\texttt{from module import X} creates a local binding to the original object that is not updated when the module variable is reassigned. Sub-patterns (c) and (d) were particularly costly to diagnose: together they required three sessions and two failed repair attempts before the root cause was identified (bugs B36--B37, B39).

\paragraph{Module bug density.}
\texttt{MLLeafPhotosynthesisMod} and \texttt{MLCanopyTurbulenceMod} each contain 8 bugs, jointly accounting for 33\% of the total (Figure~\ref{tab:bug_taxonomy}). Both modules contain iterative solvers---the photosynthesis $c_i$ scan and the Obukhov length secant method---that generate T5 gradient explosions, and both produce the highest per-sub-step arithmetic density driving T1 patterns. \texttt{MLLeafPhotosynthesisMod} additionally couples parameter tables via \texttt{MLpftcon} injection, exhibiting T2, T1, T3, and T5 bugs; \texttt{MLCanopyTurbulenceMod} exhibits T1, T3, T5, and T6. Teams translating comparable coupled land surface models should pre-audit these module classes prior to invoking \texttt{jax.grad}.

\begin{table}[h]
\centering\small
\caption{Differentiability bug taxonomy. \emph{Avg.\ att.}: mean
agent debugging iterations per bug instance. \emph{n}: total bugs of this type.
All statistics from \texttt{CHANGELOG.md} sessions 1--46
(April 1 -- May 8 2026).}
\label{tab:bug_taxonomy}
\begin{tabular}{clrrrp{5.2cm}}
\toprule
Code & Type & $n$ & \% & Avg.\ att. & Root mechanism \\ \midrule
T1 & NaN Gradient           & 17 & 35.4 & 1.0 & JAX evaluates both \texttt{jnp.where} branches; masked branch with $x{=}0$ gives $0{\times}\infty{=}\mathrm{NaN}$ \\
T2 & Zero/Wrong Gradient    &  9 & 18.8 & 1.3 & Parameter not reaching JAX trace (Python cast, state overwrite, JIT constant, local binding) \\
T3 & XLA Recompilation      &  2 &  4.2 & 1.0 & JIT cache miss per call (unstable closure, missing \texttt{lru\_cache}) \\
T4 & Memory / OOM           &  3 &  6.3 & 1.7 & Trace graph too large for device (gradient unrolling, CPU vmap unroll, large tensor) \\
T5 & Gradient Explosion     &  3 &  6.3 & 2.0 & Jacobian accumulation through $N$-iteration solver: $\lvert J\rvert^N \to \infty$ \\
T6 & Device--Host Sync      &  3 &  6.3 & 1.0 & \texttt{np.asarray(jax\_arr)} or \texttt{float()} inside hot loops forces GPU$\to$CPU copies \\
T7 & Optimization Algorithm &  4 &  8.3 & 1.5 & Adam $\beta_2$ stall, step-index arithmetic, underdetermined system \\
T8 & Crash / Compile        &  3 &  6.3 & 1.3 & XLA backend bug, GPU contention, parallel agent race condition \\
T9 & Diagnostic Reliability &  3 &  6.3 & 1.3 & FD $\varepsilon$ instability, \texttt{spval} contamination in loss, wrong timing barrier \\ \midrule
\textbf{All} & & \textbf{48} & \textbf{100} & \textbf{1.33} & 81.2\% fixed on first attempt; 18.8\% required human re-specification \\ \bottomrule
\end{tabular}
\end{table}

%% Figure: timeline
\begin{figure}[h]
  \centering
  \includegraphics[width=\linewidth]{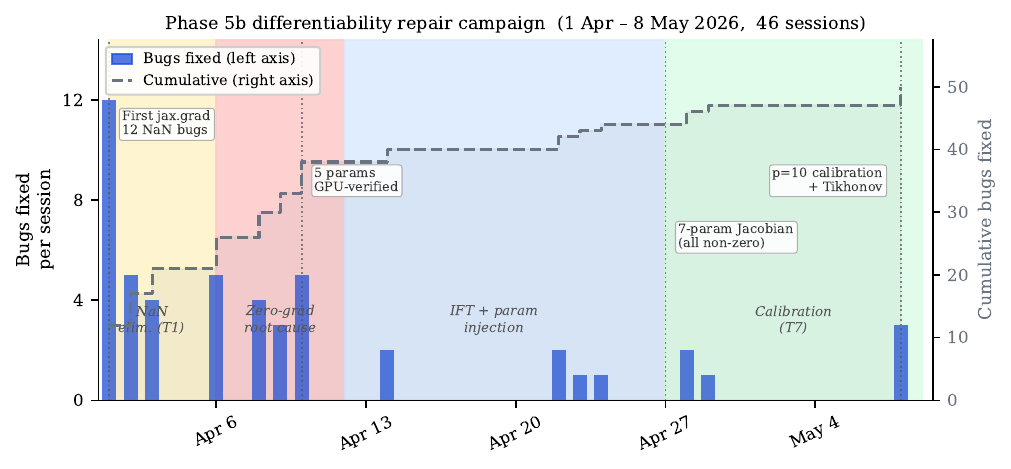}
  \caption{Differentiability repair campaign timeline
  (April 1 -- May 8, 2026, 46 sessions). Stems show bugs fixed per session.
  The dashed line shows the cumulative count (right axis). Background shading marks the
  Four dominant phases: NaN elimination (orange), zero-gradient root cause (salmon),
  IFT and parameter-injection fixes (blue), calibration and optimization (green).
  Full 7-parameter Jacobian with all non-zero columns was confirmed on April 27,
  27 days after the first \texttt{jax.grad} call.}
  \label{fig:repair_timeline}
\end{figure}

\begin{table}[h]
  \centering
  \caption{Float32 vs.\ float64 throughput on the NVIDIA A40 GPU.
           Values are amortized cost per ensemble member (ms/sample), median of 5 repeats.
           Float32 offers no measurable advantage; the model is memory-bandwidth limited.}
  \label{tab:precision}
  \begin{tabular}{rrrrr}
    \toprule
    $N$ & f64 (ms/sample) & f32 (ms/sample) & f32/f64 ratio \\
    \midrule
    1    & 29.2 & 27.6 & 0.95 \\
    8    & 13.5 & 13.5 & 1.00 \\
    32   & 11.5 & 11.9 & 1.04 \\
    128  & 11.2 & 11.5 & 1.03 \\
    512  & 11.4 & 11.6 & 1.02 \\
    1024 & 11.8 & 11.4 & 0.97 \\
    2048 & 10.9 & 11.5 & 1.05 \\
    \bottomrule
  \end{tabular}
\end{table}

\end{document}